\documentclass[letterpaper, 10 pt, conference]{ieeeconf}

\usepackage{times}
\usepackage{epsfig}
\usepackage{graphicx}
\usepackage{amsmath}
\usepackage{amssymb}
\usepackage[breaklinks=true,bookmarks=false]{hyperref}
\usepackage{subcaption}
\usepackage{comment}
\usepackage[justification=centering]{caption}
\usepackage{adjustbox}

\begin{document}

\title{An Application of Generative Adversarial Networks for Super Resolution Medical Imaging}

\author{Rewa Sood\\
{\tt\small rrsood@stanford.edu}
\and
Binit Topiwala\\
{\tt\small topbinit@stanford.edu}
\and
Karthik Choutagunta \\
{\tt\small kchoutag@stanford.edu}
\and 
Rohit Sood \\
{\tt\small rohit.sood@parexel.com}
\and 
Mirabela Rusu \\
{\tt\small mirabela.rusu@stanford.edu}
}

\maketitle

\section{Abstract}
Acquiring High Resolution (HR) Magnetic Resonance (MR) images requires the patient to remain still for long periods of time, which causes patient discomfort and increases the probability of motion induced image artifacts. A possible solution is to acquire low resolution (LR) images and to process them with the Super Resolution Generative Adversarial Network (SRGAN) to create an HR version. Acquiring LR images requires a lower scan time than acquiring HR images, which allows for higher patient comfort and scanner throughput. This work applies SRGAN to MR images of the prostate to improve the in-plane resolution by factors of 4 and 8. The term 'super resolution' in the context of this paper defines the post processing enhancement of medical images as opposed to 'high resolution' which defines native image resolution acquired during the MR acquisition phase. We also compare the SRGAN to three other models: SRCNN, SRResNet, and Sparse Representation. While the SRGAN results do not have the best Peak Signal to Noise Ratio (PSNR) or Structural Similarity (SSIM) metrics, they are the visually most similar to the original HR images, as portrayed by the Mean Opinion Score (MOS) results.

\section{Introduction}
Acquiring high-resolution (HR), clinically usable MR images is time consuming, expensive, and uncomfortable for the patient. One way to address this issue is to reduce the amount of time required to obtain each image. Outside of improving the MRI acquisition and its quality, an increase in scanner throughput can be achieved by acquiring low-resolution (LR) images instead of HR images. Ideally, these LR images can subsequently be post-processed to form super resolved (SR) images of the same perceptual quality as the original. There are various Machine Learning (ML) based solutions to the SR task, however, they have largely been applied to natural images. One of the few ML SR solutions in medical imaging is the DeepResolve network developed by Chaudhari et al~\cite{doi:10.1002/mrm.27178}, which is a 3D residual based deep learning network that achieves better results than tricubic interpolation with respect to the Peak Signal to Noise Ratio (PSNR) and Structural Similarity (SSIM) metrics. Recently, Ledig et al proposed the SRGAN which uses a perceptual loss that produces the most visually pleasing results in natural images so far~\cite{DBLP:journals/corr/LedigTHCATTWS16}. We propose to use the SRGAN to produce SR versions of LR MR images with the idea that the perceptual quality and edge fidelity of an image is more relevant to a radiologist than the PSNR or SSIM.  We expand the SRGAN implementation to work with grayscale MR prostate images at both a 4 and $8\times$ increase in resolution and compare the SR images to Sparse Representation~\cite{5466111}~\cite{10.1007/978-3-642-27413-8_47}, SRCNN~\cite{2015arXiv150100092D}, and SRResNet~\cite{DBLP:journals/corr/LedigTHCATTWS16} besides the bicubic interpolation baseline. 

\begin{table*}[!h]
    \centering
    \begin{adjustbox}{width=0.55 \textwidth,center=\textwidth}
	\begin{tabular}{| l | l | }
    \hline
    	Symbol & Definition  \\
    \hline
    $W, H$ & Width, Height of particular image \\
    $W_{i,j}, H_{i,j}$ & Width, Height of particular VGG19 feature map \\ 
    $C$ & Number of Channels \\
    $r$ & Upscaling factor \\
    $I^{LR}$ & Low-resolution image  \\
    $I^{HR}$ & High-resolution image  \\
    $I^{SR}$ & Super-resolved image \\
    $G_{\theta_G}$ & Generator Network \\
    $D_{\theta_D}$ & Discriminator Network \\
    $\theta_G$ & Generator weights and biases  \\
    $\theta_D$ & Discriminator weights and biases \\
    \hline
    \end{tabular} 
    \end{adjustbox}
    \caption{Summary of symbols and definitions}
    \label{table:symbols}
\end{table*} 

\section{Related Work}
Many of the commonly used SR techniques are non-ML based methods. In~\cite{Freeman2000}, Freeman et. al. model the relationships between low-resolution image patches and their high-resolution counterparts in a Markov network and use Bayesian belief propagation to achieve SR. In~\cite{1315043}, Chang et. al. recognized that small image patches in LR and HR images form manifolds with similar local geometry in two distinct feature spaces, and thus used manifold learning techniques such as locally linear embeddings to super-resolve images. In~\cite{5466111}, Yang et. al. use sparse coding to learn LR and HR dictionaries and use compressed sensing techniques to achieve SR. Zeyde et. al. further refined this technique by assuming a Sparse-Land model prior, which acts as a form of regularization~\cite{10.1007/978-3-642-27413-8_47}. In~\cite{5701777}, Dong et. al. use adaptive sparse domain selection and adaptive regularization to further improve image SR quality. 

In~\cite{6460827}, Gu et. al. use simple linear regression to learn mapping functions to a high-frequency manifold from different areas of middle-frequency manifolds. Combining sparse learned dictionaries, neighbor embedding methods, global collaborative coding and anchored neighborhood regression, Timofte et. al.~\cite{6751349} achieved two orders of magnitude speed improvements over the state of the art for similar or improved quality super-resolved images. In ~\cite{7299003}, Schulter et. al. use random forests to directly map from low to high-resolution image patches. One of the drawbacks of these non-ML based methods is that they require a lot of space to store the dictionaries and a significant amount of time to process the set of images. These shortcomings preclude these methods from being clinically useful. Additionally, these methods are unable to reproduce the high frequency information found in the HR images, resulting in the potential loss of important features.

The method proposed by Dong et. al. was the first deep learning architecture to solve single image SR~\cite{2015arXiv150100092D}. It used a deep convolutional neural network (CNN) to directly learn an end-to-end mapping between LR and HR images and was able to outperform sparse-coding-based methods after sufficient training. In ~\cite{doi:10.1002/mrm.27178}, Chaudhari et. al. develop a 3D CNN for medical image SR called DeepResolve to learn residual-based transformations between high-resolution thin-slice images and lower-resolution thick-slice images. While these methods perform well with respect to PSNR and SSIM, they are unable to preserve the high frequency information contained within the original HR image. The approach proposed in this paper focuses on the perceptual quality and edge fidelity of the SR image instead of the PSNR and SSIM. By using a GAN instead of a deep network and using the perceptual loss in addition to the standard losses, our method produces SR images that are visually closer to the ground truth HR image than other ML and non-ML based methods.

\section{Methods}
The following subsections discuss the notation and definitions used in this work, the image preprocessing methods, and the experimental design.

\subsection{Notation and Definitions}
Table \ref{table:symbols} contains a summary of symbols used in the rest of the paper and their definitions. Below  is a list of the equations referenced in both this section and the next.

\subsubsection{Pixel-wise Mean Squared Error (MSE) loss}
\begin{equation}\label{eq:mse_loss}
l_{MSE}^{SR}=\frac{1}{r^2WH}\sum_{x=1}^{rW}\sum_{y=1}^{rH}(I_{x,y}^{HR}-G_{\theta_G}(I^{LR})_{x,y}^2)
\end{equation}

\subsubsection{VGG loss~\cite{DBLP:journals/corr/LedigTHCATTWS16}}
\begin{equation}\label{eq:vgg_loss}
\begin{split}
l_{VGG/i,j}^{SR}=\frac{1}{W_{i,j}H_{i,j}}\sum_{x=1}^{W_{i,j}}\sum_{y=1}^{H_{i,j}}(\phi_{i,j}(I^{HR})_{x,y} \\
- \phi_{i,j}(G_{\theta_G}(I^{LR}))_{x,y})^2
\end{split}
\end{equation}
Where $\phi_{i,j}$ indicates the feature map obtained by the j-th convolution before the i-th maxpooling layer within the VGG19 network. VGG19 is a 19 layer network pretrained on the ImageNet dataset.

\subsubsection{Content loss}
The content loss is made up of the MSE and features extracted from a high-level layer in VGG19. The extracted features help in learning perceptual similarity. This network was used instead of other alternatives because it has stride-1 convolutions in the first several layers, so it retains much of the spatial information, which is important in SR.
\begin{equation}\label{eq:content_loss}
l_X^{SR} = l_{MSE}^{SR} + l_{VGG/i,j}^{SR}
\end{equation}

\subsubsection{Adversarial loss component}
The adversarial component depends on the discriminator's evaluation of the generator's output. 
\begin{equation}\label{eq:aversarial_component_loss}l_{Gen}^{SR} = \sum_{n=1}^{N}-logD_{\theta_D}(G_{\theta_G}(I^{LR}))
\end{equation}
We are minimizing $l_{Gen}^{SR}$ here because this method exhibits better gradient behavior, as opposed to minimizing $log[1-D_{\theta_D}(G_{\theta_G}(I^{LR}))]$.

\subsubsection{Perceptual loss}
Perceptual loss is a weighted sum of content loss and adversarial loss.
\begin{equation}\label{eq:perceptual_loss}
l^{SR} = l_X^{SR} + 10^{-3}l_{Gen}^{SR}
\end{equation}

\subsubsection{PSNR}
The PSNR of the SR image relative to the HR ground truth can be computed as: 
\begin{equation}\label{eq:psnr}
PSNR(I^{SR}, I^{HR}) = 10 \cdot \log_{10}\left(\frac{255^{2}}{MSE(I^{SR}, I^{HR})}\right)
\end{equation}
where 
\begin{equation}\label{eq:mse}
MSE(I^{SR}, I^{HR}) = \frac{1}{WH}\sum\limits_{i=1}^m\sum\limits_{i=1}^n \left(I^{SR}_{ij} - I^{HR}_{ij} \right)^2
\end{equation}

\begin{figure*}[!h]
    \centering
    \begin{subfigure}[!h]{0.3\textwidth}
        \centering
        \includegraphics[height=1.75in]{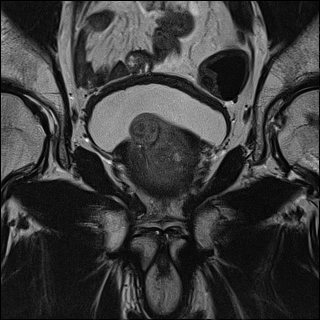}
        \caption{Ground Truth}
        \label{fig:GT}
    \end{subfigure}%
    \begin{subfigure}[!h]{0.3\linewidth}
        \centering
        \includegraphics[height=1.75in]{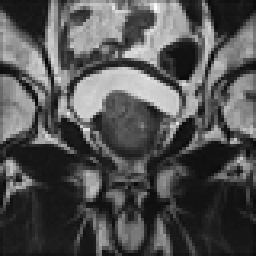}
        \caption{Low resolution input}
        \label{fig:LR}
    \end{subfigure}
    \begin{subfigure}[!h]{0.3\linewidth}
        \centering
        \includegraphics[height=1.75in]{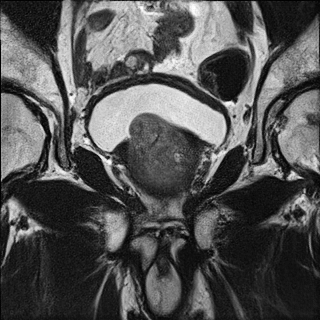}
        \caption{Proposed SR Result}
    \end{subfigure}%
    \caption{Example of Data and Proposed Result}
    \label{fig:Data}
\end{figure*}

\subsubsection{SSIM}
The SSIM is a measure that compares local patterns of pixel intensities that have been normalized for luminance and contrast~\cite{1284395}. This measure tries to account for texture changes between two images, which an MSE-based metric like PSNR cannot determine. SSIM can be written as:
\begin{equation}
\begin{split}
\label{eq:ssim} SSIM(I^{SR}, I^{HR}) = f(l(I^{SR}, I^{HR}), c(I^{SR}, I^{HR}), \\
s(I^{SR}, I^{HR}))
\end{split}
\end{equation}
where the three functions $l(I^{SR}, I^{HR})$, $c(I^{SR}, I^{HR})$, $s(I^{SR}, I^{HR})$ compare luminance, contrast and structure, respectively and $f$ is the combination function lower bounded by -1 and upper bounded by 1.

\subsubsection{MOS}
The MOS represents the perceptual quality of the image as determined by the opinion of the individual viewing the image. Specifically, an image ranked 1 would be pixelated or severely blurred while an image ranked 5 would have edge representation on par with its associated HR image. The formula for calculating the MOS is:
\begin{equation}
MOS=\frac{\sum_{n=1}^N R_n}{N}
\end{equation}
where R is the score for each image n in the set of N images.

\subsection{Pre-Processing}
We first convert the DICOM format of the input images to PNG to make them compatible with our models and resize the images so that the random cropping function, which produces $224\times 224$ crops, works correctly. We additionally scale the images so that the pixel values range from -1 to 1. An example of the data and proposed SR solution is shown in Fig. \ref{fig:Data}. We run all of the ML SR techniques on Google Cloud using an NVIDIA Tesla K80 GPU to accelerate training. 

\subsection{Experimental Design}
We use the Prostate-Diagnosis~\cite{CIAWikiDiagnosis} and PROSTATEx~\cite{CIAWikiChallenges} datasets from the Cancer Imaging Archive in this study. There are 329 patients over both sets, with multiple MR image slices per patient. These datasets contain sagittal, coronal, and/or axial prostate images, where all the prostates are affected by cancer. Data from 320 patients were included in the training set, while data from the remaining 9 were used for testing. 

We bicubicly downsample these images by a factor of 4 or 8 to form the LR input to the SRGAN. We compare the SR output of the SRGAN to the outputs of Sparse Representation, SRCNN, and SRResnet, using bicubic interpolation as a baseline. The images produced by the Sparse Representation algorithm provide insight into how well non-ML based algorithms accomplish the SR task on medical images. SRCNN and SRResNet provide understanding on how a shallow and a deep ML network accomplish this same task. We discuss each method further below.

\begin{figure*}[!h]
    \centering
    \begin{subfigure}[!h]{0.5\textwidth}
        \centering
        \includegraphics[height=1.5in]{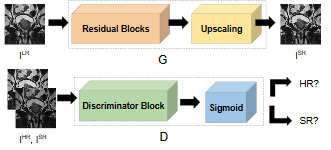}
        \caption{SRGAN Model}
        \label{fig:srgan-architecture}
    \end{subfigure}%
    \begin{subfigure}[!h]{0.45\linewidth}
        \centering
        \includegraphics[height=1.3in]{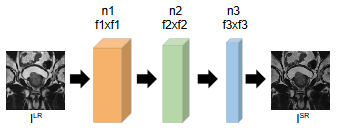}
        \caption{SRCNN Model. $(f_1, f_2, f_3) = (9, 1, 5)$ - filter shape. $(n_1, n_2, n_3) = (64, 32, 1)$ - number of filters.}
        \label{fig:srcnn-architecture}
    \end{subfigure}
    \caption{Model architectures}
    \label{fig:Architectures}
\end{figure*}

\subsection{SRGAN}
SRGAN is the first framework capable of inferring photo-realistic natural images for $4\times$ upscaling factors~\cite{DBLP:journals/corr/LedigTHCATTWS16}. The architecture in Fig. \ref{fig:srgan-architecture} consists of two parts: the generator and the discriminator networks. The generative model is a deep residual network that accepts an LR image and outputs an SR image. The advantage of having a residual network with skip connections is that the generator avoids vanishing and exploding gradients, which could arise due to the depth of the network. The generator is trained with the goal of fooling the discriminator into believing that the output SR images are HR. The discriminator is trained to distinguish SR images from the original images. The GAN approach uses a loss function that is comprised of a perceptual loss (eq. \ref{eq:perceptual_loss}), which encourages SR reconstructions to move towards regions of the search space with high probability of containing photo-realistic images and a content loss (eq. \ref{eq:content_loss}) based on perceptual similarity using the high-level features from a pretrained VGG19 network~\cite{DBLP:journals/corr/JohnsonAL16}.

The original network works with RGB images, while MR images are grayscale. We modify the network to work with a one channel input, however, the VGG19 pretrained network requires three channels and cannot be modified. To accommodate this constraint, we replicate the grayscale images to form three channels only in the step that calculates the perceptual loss. Initially we execute the SRGAN network over 500 randomly chosen training images using an Adam optimizer with a batch size of 16, learning rate of 0.0001 and beta1 as 0.9 to ensure that the SRGAN was generating an SR image comparable to the original HR image. We then train the model on the full training dataset with the same hyperparameters. The generator was executed standalone for 20 epochs and then the discriminator and generator were trained sequentially for 50 epochs. The generator has the difficult task of generating a new image, while the discriminator only has to solve a simple classification problem. Thus, training the generator network on its own for 20 epochs ensures that the discriminator will not dominate the learning process and prevent the generator from learning anything at all.

\subsection{SRCNN}
SRCNN~\cite{2015arXiv150100092D} has three convolutional layers (Fig. \ref{fig:srcnn-architecture}). The first layer learns feature maps from the LR images. The second layer can be intuitively thought of as learning feature maps for the SR image from the LR feature maps. The final layer uses these SR feature maps to construct the actual SR image. The ReLU non-linearity is applied to the two initial layers. The first layer uses 64 filters of size 9x9, the second layer uses 32 filters of size 1x1, and the last layer uses a filter of size 5x5. The SRCNN uses the MSE (eq. \ref{eq:mse_loss}) as its loss function. We first train the SRCNN with a subset of 500 random training images for 100 epochs with a learning rate of 0.0001 and batch size of 128 with the Adam optimizer. We then train the SRCNN on the full training dataset for 500 epochs, using the same hyperparameters.

\subsection{SRResNet}
SRResNet is the generator component of the SRGAN network. We first overfit this network to 500 randomly chosen training images to ensure that we correctly extract the generator portion of the network. We then run the full training subset for 50 epochs with a learning rate of 0.0001 and batch size is 16, giving over 875 iterations per epoch. The loss optimized via Adam is the sum of the pixel-wise MSE loss (eq. \ref{eq:mse_loss}) and the VGG loss (eq. \ref{eq:vgg_loss}) for perceptual similarity, and contained no generator or discriminator loss.

\subsection{Sparse Representation}
Sparse representation is a non-deep learning approach to single image super resolution ~\cite{5466111}. The underlying idea behind this method is the observation that most image patches can be well-represented by a sparse linear combination of elements from a well-chosen over-complete dictionary. Recent results from compressed sensing literature suggest that it is possible to correctly recover a sparse representation of an LR input image and use the coefficients of this representation to generate an SR output image. The sparse representation algorithm first trains two dictionaries for the HR and LR images. Then each input LR image is divided into smaller patches. For each patch of a given LR input image, the algorithm reconstructs an HR image patch by solving an optimization problem with sparsity constraints. Finally, the algorithm pieces together various HR patches to form a single SR image by solving another convex optimization problem via gradient descent. The loss function used in the optimization contains terms for local consistency and global consistency, so that the generated high-resolution image looks like a single, coherent image. We modify the MATLAB implementation provided in~\cite{githubSparseRep} for MR images. The results presented in this paper use the following hyperparameters: sparsity regularization $\lambda = 0.2$, patch sizes of $5\times 5$ pixels, and a maximum of 20 iterations for the backprojection step. Because the training procedure for sparse representation requires creating dictionaries of LR and HR image patches which can be quite space intensive to store on disk, we are limited to training the dictionaries with only 36 example MRI images. During the training process, one thousand random samples of $5\times 5$ pixel patches were obtained from each image to form the dictionaries. During evaluation, each LR test image was upscaled by $2\times$ twice to create a resulting SR image with an effective upscaling factor of $4\times$.

\section{Results}
In the first two subsections we discuss some experiments run on SRCNN and SRResNet and their results. The last two subsections discuss the qualitative and quantitative comparison between the SRGAN SR output and the SR output of the rest of the models. 

\begin{figure}[h!]
	\centering
      \includegraphics[height=1.8in]{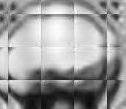}
      \caption{artifacts in SRCNN}
    \label{fig:srcnn-artifacts}
\end{figure}

\begin{figure*}[h!]
    \centering
    \begin{subfigure}[h!]{0.3\textwidth}
        \centering
        \includegraphics[height=1.75in]{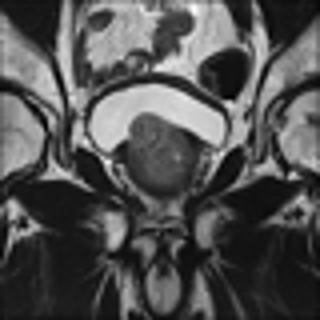}
        \caption{Bicubic interpolation}
    \end{subfigure}%
    \begin{subfigure}[h!]{0.3\textwidth}
        \centering
        \includegraphics[height=1.75in]{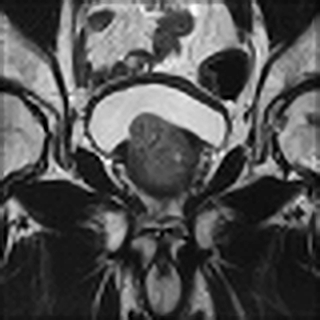}
        \caption{Sparse Representation}
        \label{fig:sparse-rep-result-good}
    \end{subfigure}%
    \begin{subfigure}[h!]{0.3\textwidth}
        \centering
        \includegraphics[height=1.75in]{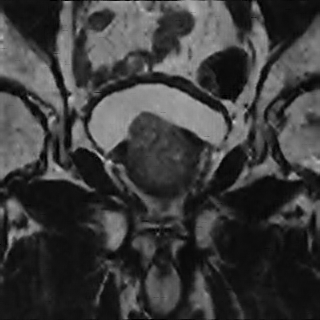}
        \caption{SRCNN}
    \end{subfigure}

    \begin{subfigure}[h!]{0.3\textwidth}
        \centering
        \includegraphics[height=1.75in]{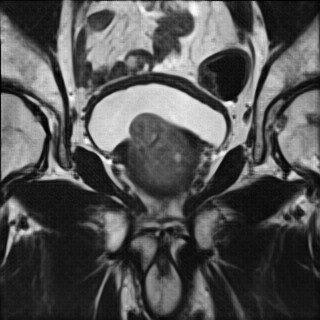}
        \caption{SRResNet}
    \end{subfigure}%
    \begin{subfigure}[h!]{0.3\textwidth}
        \centering
        \includegraphics[height=1.75in]{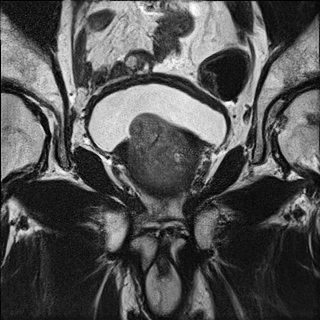}
        \caption{SRGAN 4x}
    \end{subfigure}%
    \begin{subfigure}[h!]{0.3\textwidth}
        \centering
        \includegraphics[height=1.75in]{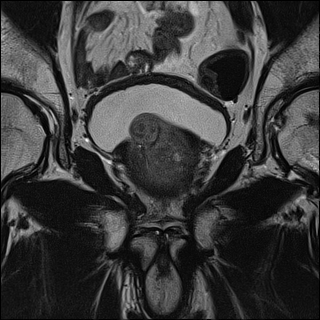}
        \caption{HR ground truth image}
    \end{subfigure}
    \caption{SR result for $4\times$ upscaling using- Bicubic interpolation, SRCNN, Sparse Representation, SRRestNet, and SRGAN}
    \label{fig:SRGANComparison}
\end{figure*}

\begin{figure*}
\centering
    \begin{subfigure}[!h]{0.30\textwidth}
    	\centering
        \includegraphics[height=1.75in]{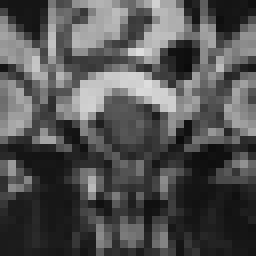}
        \caption{LR input to SRGAN 8x}
    \end{subfigure}
    \begin{subfigure}[!h]{0.30\textwidth}
        \centering
        \includegraphics[height=1.75in]{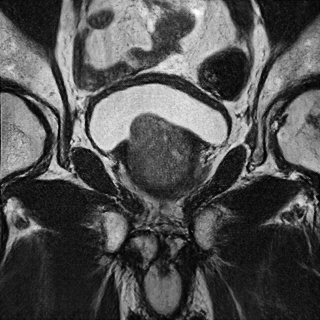}
        \caption{SRGAN 8x}
    \end{subfigure}%
    \caption{SRGAN result for $8\times$ upscaling compared to LR input}
    \label{fig:SRGAN8x}
\end{figure*}

\subsection{SRCNN}
The vanilla SRCNN uses a normal distribution with defined standard deviation for initializing filter weights. The resulting SR image does not look any different from the input LR image after training for a few epochs. We replace this weight initialization method with the Kaiming He initialization~\cite{DBLP:journals/corr/HeZR015} and the model starts showing improvements in the generated SR images. The SRCNN only performs convolution operations without padding. We try the following strategies to ensure the SR image has the same shape as the original HR image: modify the SRCNN architecture to pad the input to each convolution operation and retain the existing SRCNN architecture and pad (zero or reflective) the input LR image. When we train the model with the first approach, the resulting SR image has gridding artifacts (Fig. \ref{fig:srcnn-artifacts}). The second approach with zero padding also produces artifacts, however the same approach with reflective padding produces no artifact. It seems that the SRCNN model is sensitive to the sharp edges produced by zero padding, which are not as prominent with reflective padding.

\subsection{SRResNet}
The images acquired via MRI follow a protocol and are generally somewhat similar among different patients. Given that the network has many comparable images to learn from, we postulate that it would have enough information to overcome a heavier penalization to produce even more accurate images. We first scaled the MSE loss by a factor of 5. As expected, this makes the resulting images blurrier. More interestingly, we train the SRResNet with two different scales for the VGG loss: the original scale and a factor of 10 increase. Unfortunately, this network produces gridding patterns with the more heavily weighted VGG loss. Apparently, a higher weighting to the VGG loss emphasizes edges to the extent that the network produces spurious edges in the resulting SR image.   

\begin{table*}[!h]
	\centering
    \begin{adjustbox}{width=0.70 \textwidth,center=\textwidth}
	\begin{tabular}{| l | l | l | l | l | l |}
    \hline
    	& Bicubic & Sparse Rep. & SRCNN & SRResNet & SRGAN  \\
    \hline
    PSNR [dB] & 21.68 & 21.82 & 24.02 & 21.03 & 21.27 \\
    SSIM & 0.71 & 0.74 & 0.68 & 0.70 & 0.66 \\
    MOS & 2.6 & 2 & 2.6 & 4 & 5 \\
    \hline
    \end{tabular}
    \end{adjustbox}
    \caption{4x SR performance results averaged across a subset of test images}
    \label{table:performance_results}
\end{table*}

\subsection{Qualitative SRGAN Image Quality Assessment}
Fig. \ref{fig:SRGANComparison} contains an example SR output for each method and the LR and HR images for reference. The LR image is severely pixelated and has no edge fidelity. While the image produced via bicubic interpolation has no pixelation, this method is still unable to preserve the high frequency information found in the ground truth image. The Sparse Representation method produces slightly better results than the previously discussed methods. However, the large amount of space and time overhead required by this model precludes it from clinical use. The SRCNN begins to show edge preservation, however the features within different regions of the output SR image are smoothed out. The SRCNN is especially biased toward smoothing the image because the network only uses MSE loss. The SRResNet has both MSE and perceptual loss yet fails to outperform the SRGAN. Clearly, the discriminator network seeks out the high frequency information that differentiates HR and LR images, thus forcing the SRGAN output to have far more high frequency details than the output of the SRResNet. The SRGAN 8x network is not able to maintain as high an edge fidelity as the SRGAN 4x network. This result is expected because the SRGAN 8x network is provided with far less information since the input LR image is a further 2x smaller in both dimensions (Fig. \ref{fig:SRGAN8x}). Overall, in comparing the SRGAN to the other models, the outputs from the SRGAN are visually closer to the original HR ground truth images.

\subsection{Quantitative SRGAN Image Quality Assessment}

For this work, we gave a radiologist five sets containing seven images each and asked them to rank each image between 1 and 5 inclusive, with 1 being bad quality and 5 being excellent quality. The radiologist was instructed to evaluate the quality of the images with respect to the edge fidelity of the image. The seven images in each set included the LR and HR images, as well as the SR outputs from Bicubic Interpolation, Sparse Representation, SRCNN, SRResnet, and SRGAN 4x. 

In Table ~\ref{table:performance_results}, we report the performance results for each method in terms of PSNR, SSIM, and MOS for SRGAN. SRGAN has one of the lowest PSNR and SSIM results, however this result is also seen in the referenced SRGAN paper \cite{DBLP:journals/corr/LedigTHCATTWS16}. Since SRGAN emphasizes edges while the other methods tend to produce smooth output images, it has a relatively high MSE, leading to a lower PSNR. Both the PSNR and SSIM metrics do not provide insight regarding image quality with respect to human perception. The MOS results, however, do reflect human perception and show that the SRGAN performs the best overall, in this respect. 

\section{Conclusion}
The SRGAN output images contain more high frequency information than the other SR approaches, which makes the SRGAN results visually closer to the original HR images. This fact is reflected in the higher average MOS that the SRGAN images receive. While the other images remove the pixelation found in the LR image, they tend to blur the image due to their emphasis on the MSE loss. By using the perceptual loss and the GAN format of training, the SRGAN does not smooth the image while still removing pixelation. Overall, the SRGAN produces the most visually pleasing results compared to output images from bicubic interpolation, SRCNN, SRResNet, and Sparse Representation.

\begin{equation}l_{D}^{SR} =-\sum_{n=1}^{N}logD_{\theta_D}(I^{LR})- \sum_{n=1}^{N}log(1-D_{\theta_D}(G_{\theta_G}(I^{LR})))
\end{equation}

{\small
\bibliographystyle{ieee}
\bibliography{egbib}
}

\end{document}